\documentclass{llncs}
\usepackage{amstex}
\usepackage{float}
\usepackage{psfrag}
\usepackage{graphicx}
\begin{document}
\pagestyle{headings}  
\title{A 1d Traffic Model with Threshold Parameters}
\titlerunning{1d Traffic Model}  
\author{G. Sauermann \and H. J. Herrmann}
\authorrunning{G. Sauermann}   
%
\tocauthor{G. Sauermann, H. J. Herrmann (University of Stuttgart)}
\institute{University of Stuttgart, Institute for Computer
  Applications 1,\\ Pfaffenwaldring 27, 70569 Stuttgart, Germany
}

\maketitle              

\begin{abstract}
The basic properties of traffic flow are analyzed using a simple
deterministic one dimensional {\em car following model} with continuous
variables based on a model introduced by Nagel and Herrmann
[Physica A {\bf 199} 254--269 (1993)] including a few
modifications. As a first case we 
investigate the creation and propagation of jams in a platoon
generated by a slow leading vehicle. In a second case we look at a
system with the size $L$, periodic boundary conditions and identical
vehicles. A strong dependence on the initial
configuration of the fundamental diagram's shape can be found.
\end{abstract}

\section{Definition of the Model}

To get a parallel update of all vehicles, the first step is to change
the velocity of all vehicles taking into account the threshold
parameters $\alpha$ and $\beta$ according to the rules defined below. In
the second step the position of all vehicles is changed using the
velocity calculated in the first step.

Position, velocity and threshold parameters of the vehicles are
continuous variables. $\Delta t$ is the time step used in the
simulations.

\paragraph{\bf 1st step - velocity:}
A vehicle decelerates if the headway distance $\Delta x_i(t)$ is
smaller than a safety distance $\alpha$. The headway distance 
after the deceleration step is determined by $\delta$.
\begin{equation}
  \Delta x_i(t) - v_i(t) \Delta t < \alpha 
  \leadsto v_i(t+\Delta t) = 
    \max\left(0, \frac{\Delta x_i(t)-\delta}{\Delta t}\right)
\end{equation}
The acceleration of a vehicle depends on a threshold parameter $\beta$
and the maximum velocity $v_{max}$:
\begin{equation}
  \Delta x_i(t) - v_i(t) \Delta t > \beta
  \leadsto v_i (t+\Delta t) = \min\left( v_{max}, v_i(t) + a \Delta t \right)
\end{equation}
The acceleration coefficient is determined by the headway distance if
$\gamma$ is greater than $\beta$ and the headway distance is smaller
than $\gamma$:
\begin{equation}
  a = a_{max} \max\left(1, \frac{\Delta x_i(t)}{\gamma}\right).
\end{equation}

\paragraph{\bf 2nd step - position:} The positions are changed.

\begin{equation}
  x_i(t + \Delta t) = x_i(t) + v(t + \Delta t) \Delta t
\end{equation}

Using a constant acceleration coefficient ($\gamma<\beta$) we can
compare this model to the deterministic case of the
Nagel-Schreckenberg model \cite{nagschreck}, as the velocity is
determined by de- and acceleration with the acceleration causing a
discretisation of the velocity.

\section{Platoon}

\begin{figure}[ht]
  \parbox{5.6cm}
  {
    \centering
    \psfrag{time}[t][]{time in s}
    \psfrag{position}[][]{position in km}
    \includegraphics[width=5.5cm]{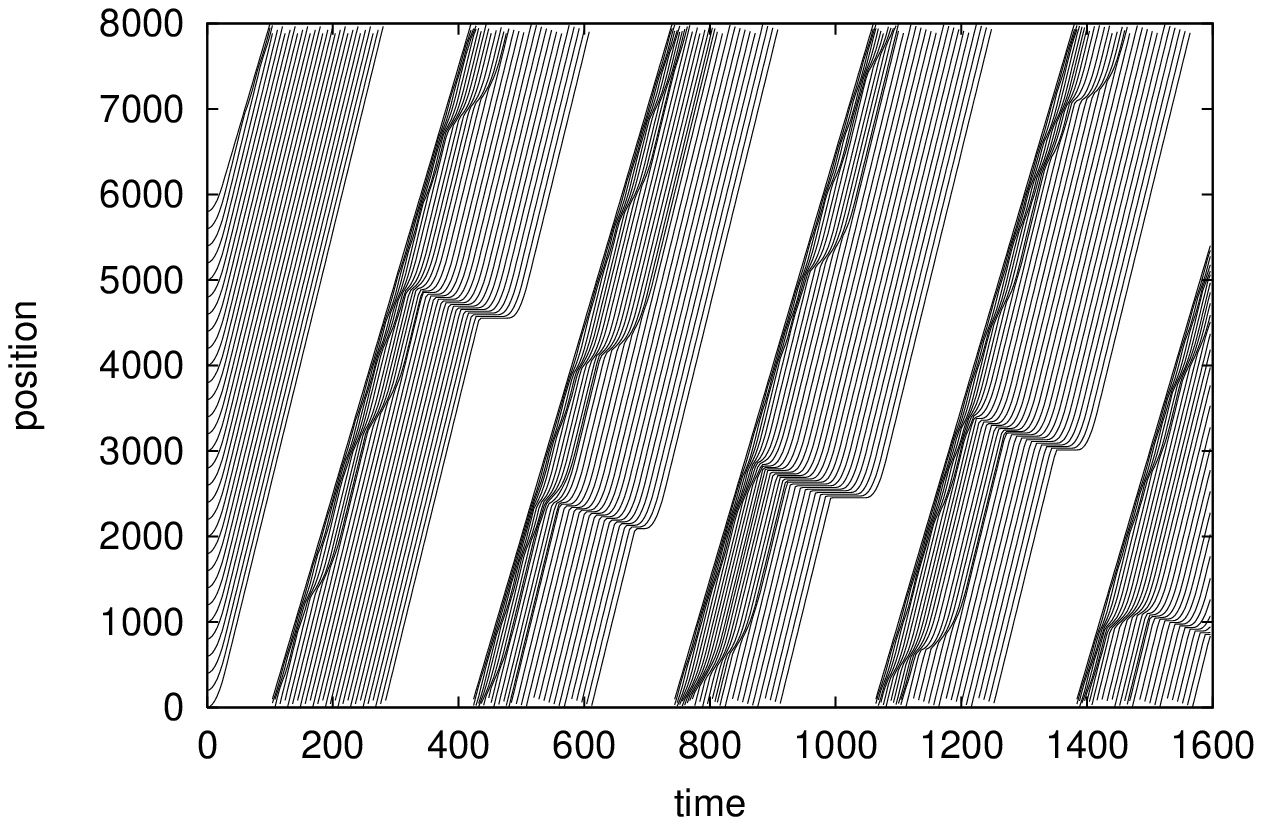}
    \medskip
    \psfrag{headway}[][]{relative position in km}
    \includegraphics[width=5.5cm]{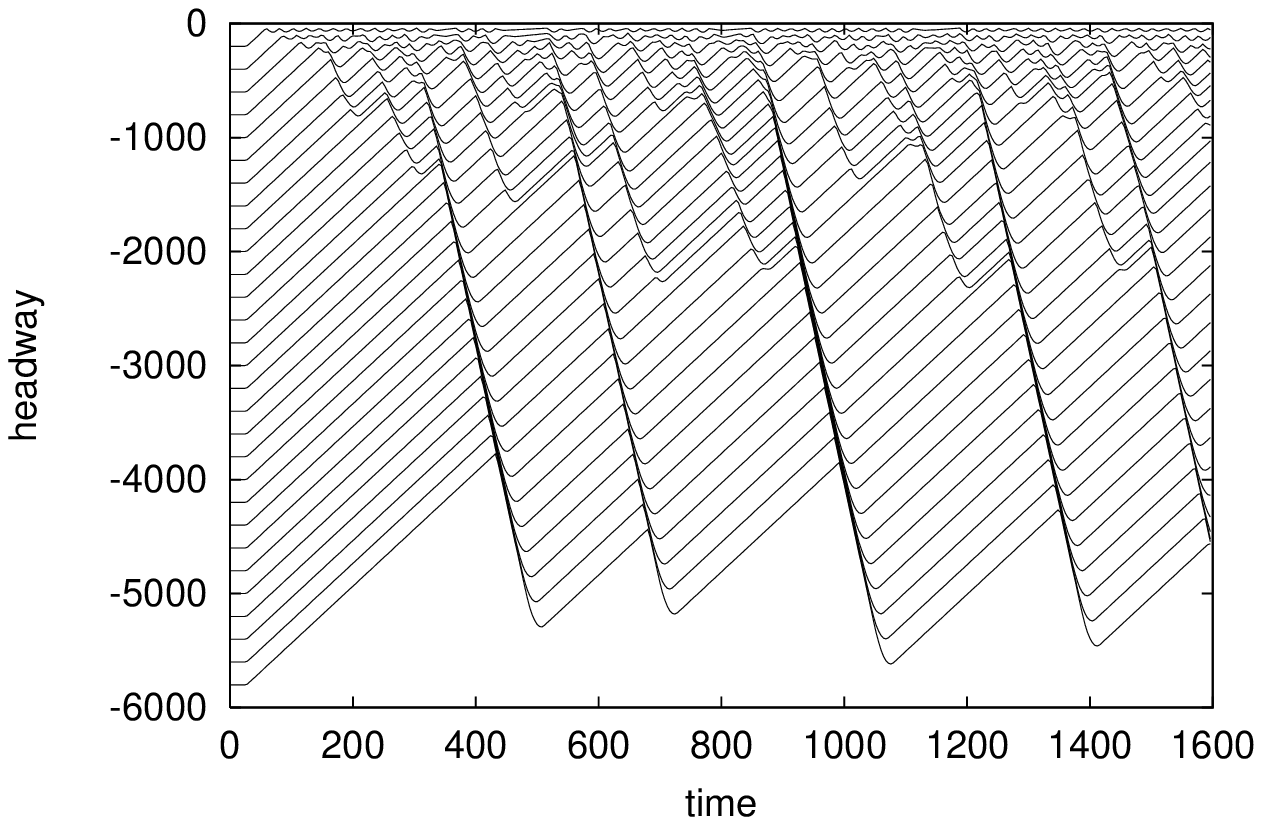}
  }
  \parbox{6.4cm}
  {
    \centering
    \psfrag{time}[t][]{time in s}
    \psfrag{distance}[][]{relative position in km}
    \psfrag{velocity}[][]{velocity in m/s}
    \psfrag{v_lead}[r][r]{\tiny $v_{lead}$}
    \psfrag{v_lead-1}[r][r]{\tiny $v_{lead-1}$}
    \psfrag{v_lead-2}[r][r]{\tiny $v_{lead-2}$}
    \psfrag{dx}[r][r]{\tiny $x_i - x_{lead}$}
    \psfrag{dx_min}[r][r]{\tiny $\Delta x_{min}$}
    \psfrag{dx_acc}[r][r]{\tiny $\Delta x_{acc}$}
    \includegraphics[width=6.3cm, height=7.8cm]{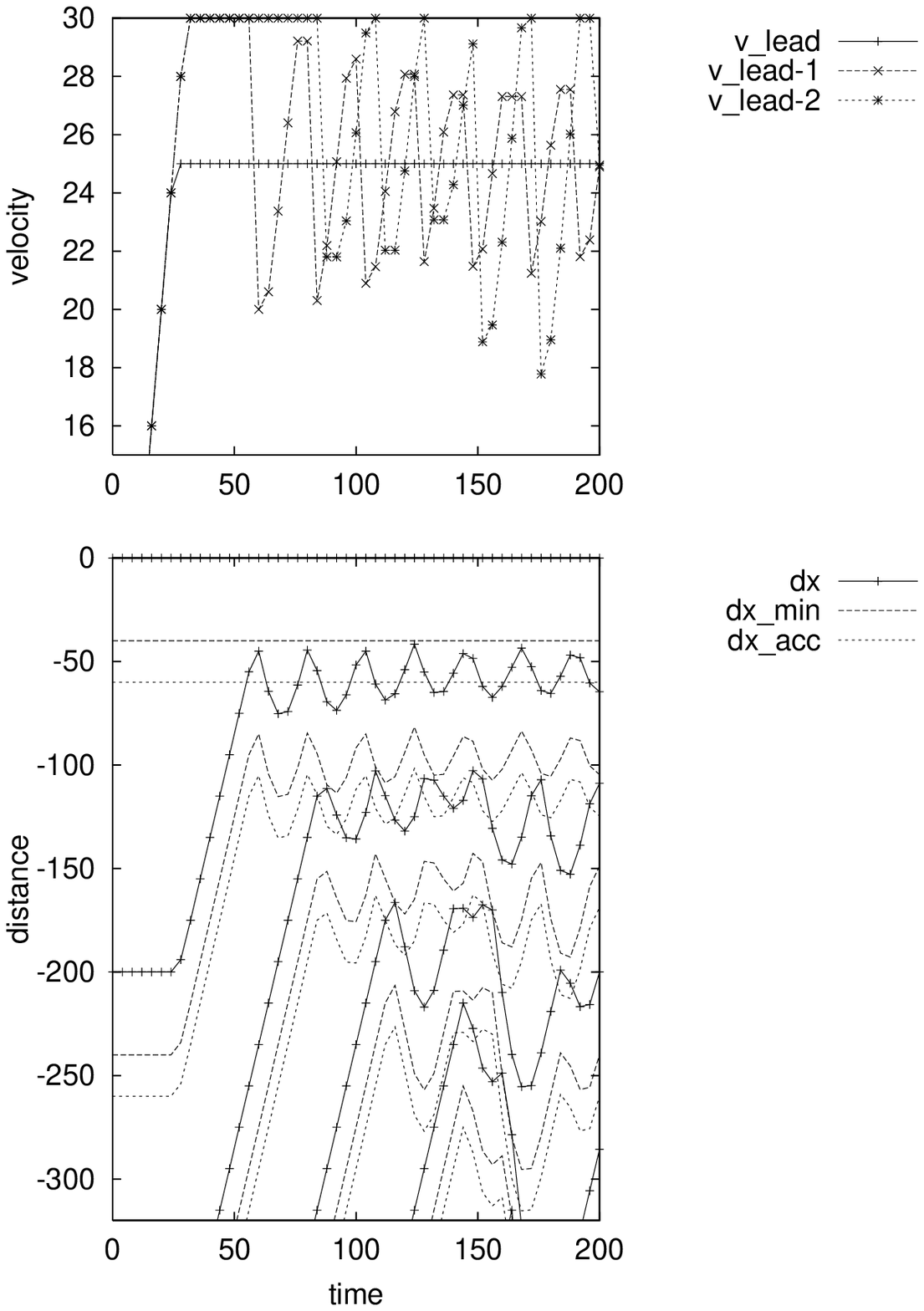}
    \medskip
  }
  \caption{The figures on the left side show the trajectories of
    vehicles and jams of different sizes. On the right side the
    oscillations of the velocity and relative position (position in
    the coordinate system of the leading vehicle) of the second
    and third vehicle are shown including the threshold conditions for
    accelerating and breaking. Parameters: $\alpha=15, \beta=35,
    \gamma=100, \delta=20, a_{max}=1, v_{max}=30, v_{lead}=25$}
\end{figure}

A slow leading vehicle that is followed by faster ones creates a platoon of
vehicles. In this platoon, jams are generated directly behind the leading
vehicle. They propagate either through the entire platoon or at least through parts
of it. These jams are created because a vehicle is not able
to slow down exactly to the velocity of the one ahead. In our
model this is forced by the parameter $\delta$, that is the headway
distance after a breaking step. For $\delta>\alpha$ the headway
distance $\alpha$ at the beginning of the breaking step is smaller
than the headway distance $\delta$ after the breaking. The consequence
is, that the vehicle has to increase the headway distance
within the breaking time step by decelerating to a slower velocity than
the car ahead. This leads to an oscillation in the velocities and the headway
distances of the vehicles and can add up in a constructive way
generating jams with the size of the entire platoon.

\section{Periodic Boundary Conditions}

In our study, we work on the fundamental diagram (flux vs. density) of a system with
the size $L$ and periodic boundary conditions using identical
vehicles. Strong dependencies on the fluctuations in the
initial configuration have been found and investigations of three
different cases have been made.
On one side we look at a highly symmetric case without
any fluctuations using equidistant start positions. On the other
side we use homogeneous random start positions.
Between these two cases, random fluctuations are used to move the
vehicles out of their equidistant positions. The maximum of these
fluctuations, however, is controlled by a parameter $\Delta L$ that
can be in the range from zero up to the system size.
The idea is that with the parameter $\Delta L$ the fluctuations
or the disorder of the initial state can be characterized including the
two extreme cases mentioned above representing the configurations
for the minimum and maximum values of $\Delta L$. For these extreme
cases analytical solutions can also be found. The velocity of all
vehicles is set to zero in the initial configuration and the
same value is used for the parameters $\alpha$ and $\delta$.

\subsection{Equidistant Starting Positions}

\begin{figure}
  \centering
  \psfrag{density}[t][]{Density $\rho$ in $vehicles/km$}
  \psfrag{flux}[][]{Flux $\Phi$ in $vehicles/h$}
  \psfrag{deltat01s}[r][r]{\tiny $\Delta t=0.1s$}
  \psfrag{deltat1s}[r][r]{\tiny $\Delta t=1s$}
  \includegraphics[width=6cm]{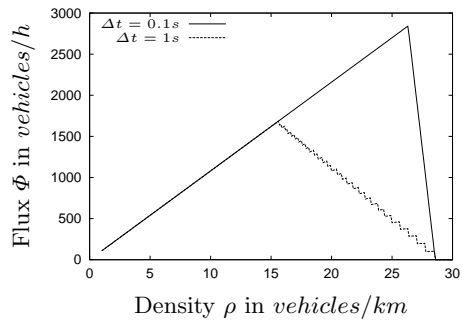}
  \caption[]{Flux for different simulation time-steps $\Delta t$ and equidistant
    start positions. Parameters: $\alpha=\delta=15, \beta=35,
    \gamma=10, a_{max}=1, v_{max}=30$}
\end{figure}

In this highly ordered case the model generates no jams and the
fundamental diagram is a straight line with a slope of $v_{max}$. The
maximum flux can be obtained from the accelerating condition of the
model.
\begin{equation}
  \rho_{crit} = \frac{1}{\Delta t v_{max} + \beta}
\end{equation}
\begin{equation}
  \Phi(\rho_{crit}) = v_{max} \rho_{crit}
\end{equation}
For $\Delta t \longrightarrow 0$ the maximum flux jumps to zero at the
maximum density $\rho_{max}=1/\beta$, which is determined only by the acceleration
threshold parameter $\beta$.

\subsection{Homogeneous Random Starting Positions}

\begin{figure}[ht]
  \parbox{6cm}
  {
    \centering
    \psfrag{time}[t][]{Time $t$ in $min$}
    \psfrag{position}[][]{Position $x$ in $km$}
    \includegraphics[width=5.5cm]{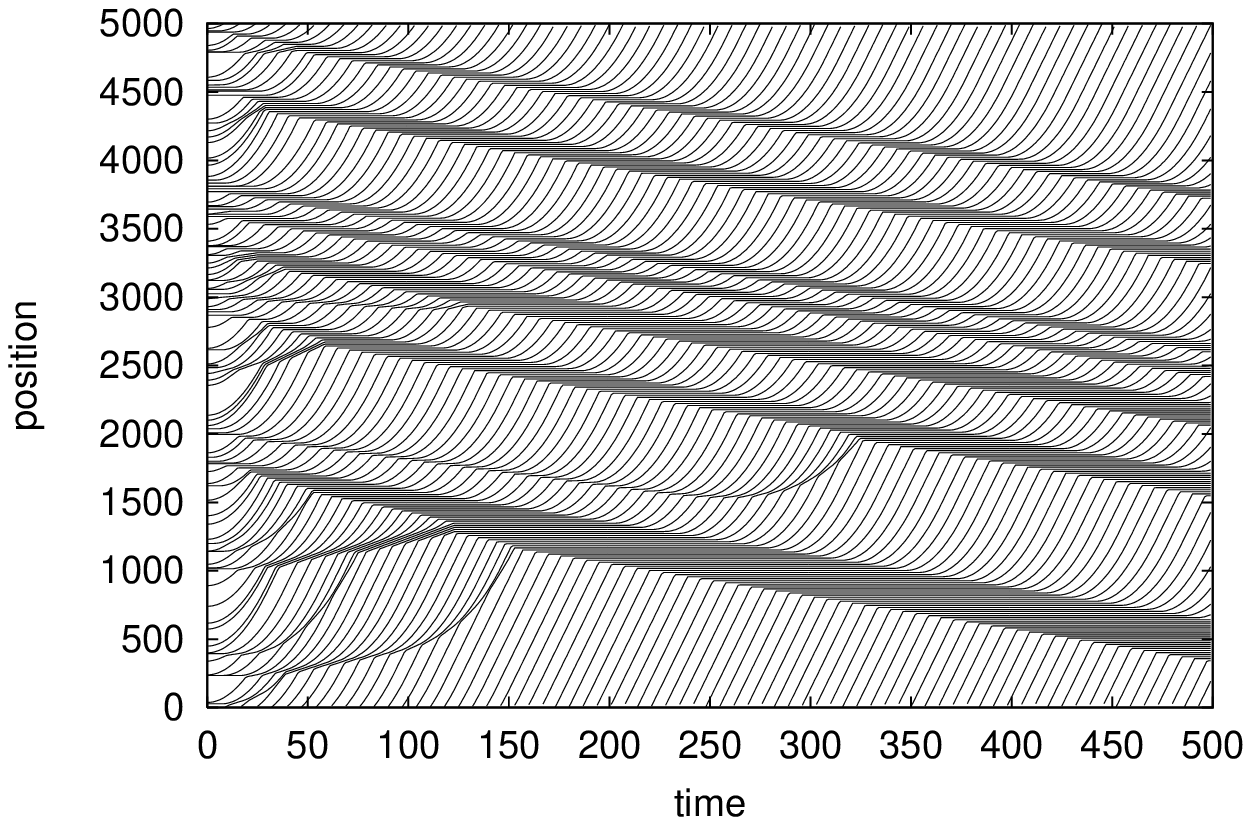}
  }
  \parbox{6cm}
  {
    \centering
    \psfrag{simulation}[r][r]{\tiny Simulation}
    \psfrag{average}[r][r]{\tiny Average}
    \psfrag{density}[t][]{Density $\rho$ in $vehicles/km$}
    \psfrag{flux}[][]{Flux $\Phi$ in $vehicles/h$}
    \includegraphics[width=5.5cm]{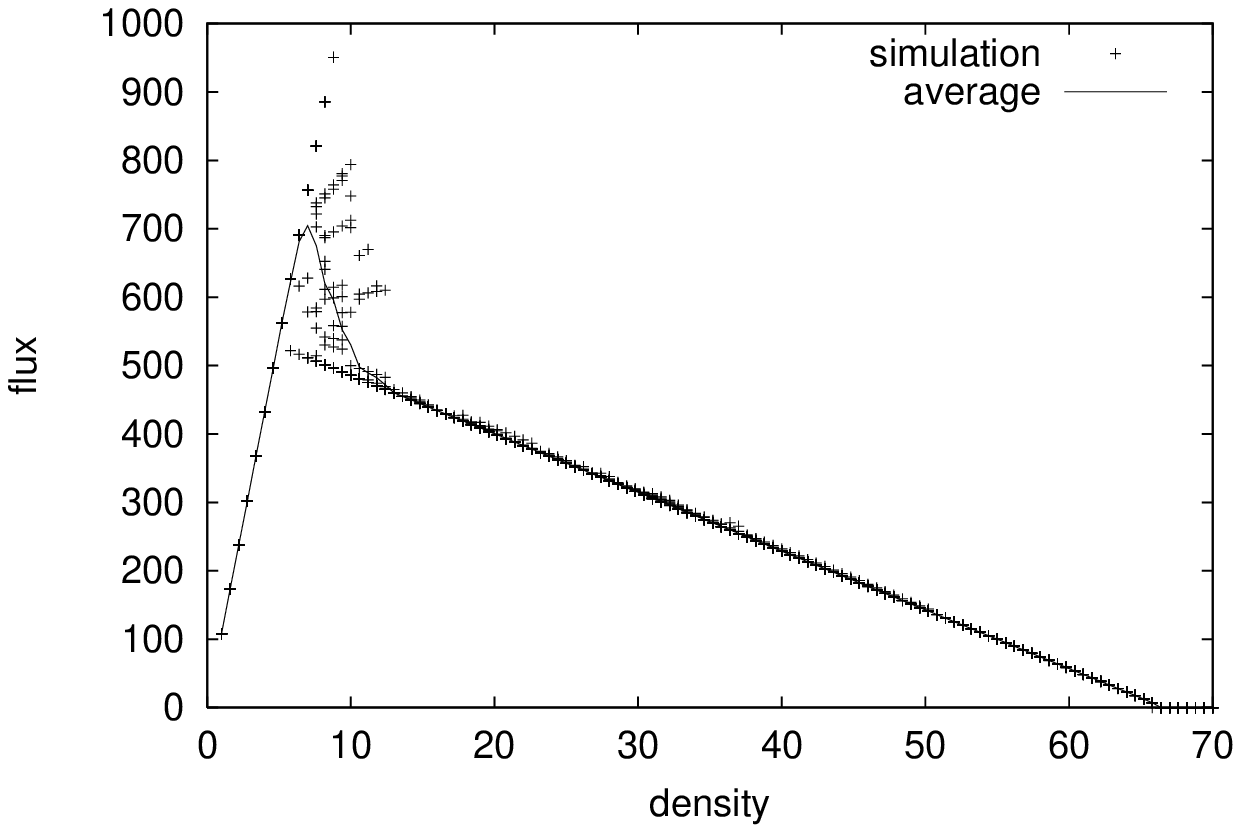}
  }
  \medskip
  \label{abs_flux}
  \caption[]{The left side shows density waves in a space time
    diagram. The right side shows the flux
    for random initial positions and various safety distances. 
    Parameters: $\alpha=\delta=15, \beta=35,
    \gamma=10, a_{max}=1, v_{max}=30$}
\end{figure}

In this case we find a triangular-shaped fundamental diagram with a
maximum density $\rho_{max}=1/\alpha$ and a critical density
$\rho_{crit}$ with maximum flux $\Phi_{max}$. Up to the critical
density the system is free-flowing and from the critical density up to the
maximum density there are both, jammed and free flowing
areas. Because of the sharp boundaries between congested and free
areas, the system can be seen as a mixture of two phases like water and
ice. The size of the jammed and the free phase, respectively, can
be obtained through the normalized (divided by the maximum flux) falling
straight line of the fundamental diagram.

Looking at the trajectories of vehicles entering a congested area, it
can be proven that the density of the congested area depends on the
velocity of the entire jammed area itself. In the final state of a
finite system, only the congested areas with the slowest velocity
survive. Taking this into account and neglecting the acceleration and
deceleration stripes, there are two phases characterized by the density
and the velocity of the vehicles within each phase. In order to obtain
the fundamental diagram for this two-phase system the mean velocity is
calculated.
\begin{equation}
  \overline{v(\rho)}
  = \frac{1}{N} \sum_{i=1}^N v_i
  = \frac{1}{N} \left( N_j v_j + N_f v_f \right)
  = \frac{v_f - v_j}{\rho_j - \rho_f}
       \left(\frac{\rho_f(\rho_j - \rho)}{\rho}\right) + v_j
  \label{e1}
\end{equation}
The indices $j$ and $f$ are used for {\em jammed} and {\em free} phase,
respectively.

The density of the congested area can be obtained from the braking rule
of the model.
\begin{equation}
  \rho_j = \frac{1}{\alpha + v_j \Delta t}
  \label{e2}
\end{equation}
The outflow of a congested area determines the density of the free
flowing area that is equal to the critical density. 
As a result of analyzing the trajectories, it is found that the time
$\Delta T$, that is needed to accelerate and pass the window
$\beta-\alpha$, is an important value in this model.
\begin{equation}
  \Delta T = \left( \left \lfloor - \frac{1}{2} + \sqrt{
    \frac{1}{4} + \frac{2(\beta-\alpha)}{a \Delta t^2}
  } \right \rfloor + 1 \right) \Delta t
  \label{e3}
\end{equation}
The braces $\lfloor .. \rfloor$ operate on the given argument by
rounding down to the next integer value.
The result for the critical density using $v_{max}=v_f$ is therefore:
\begin{equation}
  \rho_{crit} = \rho_f = \frac{1}{\alpha + v_j \Delta t +
    (v_{max}-v_j) \Delta T}
  \label{e4}
\end{equation}
From equation (\ref{e1}, \ref{e2}, \ref{e3}, \ref{e4}) the exact
solution for the flux using a constant acceleration coefficient can be obtained:
\begin{equation}
  \label{phi_rho}
  \Phi(\rho) = \overline{v}(\rho) \rho = 
    \frac{1}{\Delta T} +
    \rho \left(
      v_j - \frac{\alpha + v_j \Delta t}{\Delta T}
    \right)
\end{equation}
Investigations on the connection between the jam velocity and the
slope of the fundamental diagram have shown an exact agreement with
the results of Lighthill and Whitham \cite{ligwhit}.
\begin{equation}
  v_{jam} = \frac{d \Phi}{d \rho}
\end{equation}

\subsection{Equidistant Starting Positions with Random Fluctuation}

\begin{figure}[ht]
  \parbox{6cm}
  {
    \centering
    \psfrag{time}[t][]{Time $t$ in $min$}
    \psfrag{position}[][]{Position $x$ in $km$}
    \includegraphics[width=5.5cm]{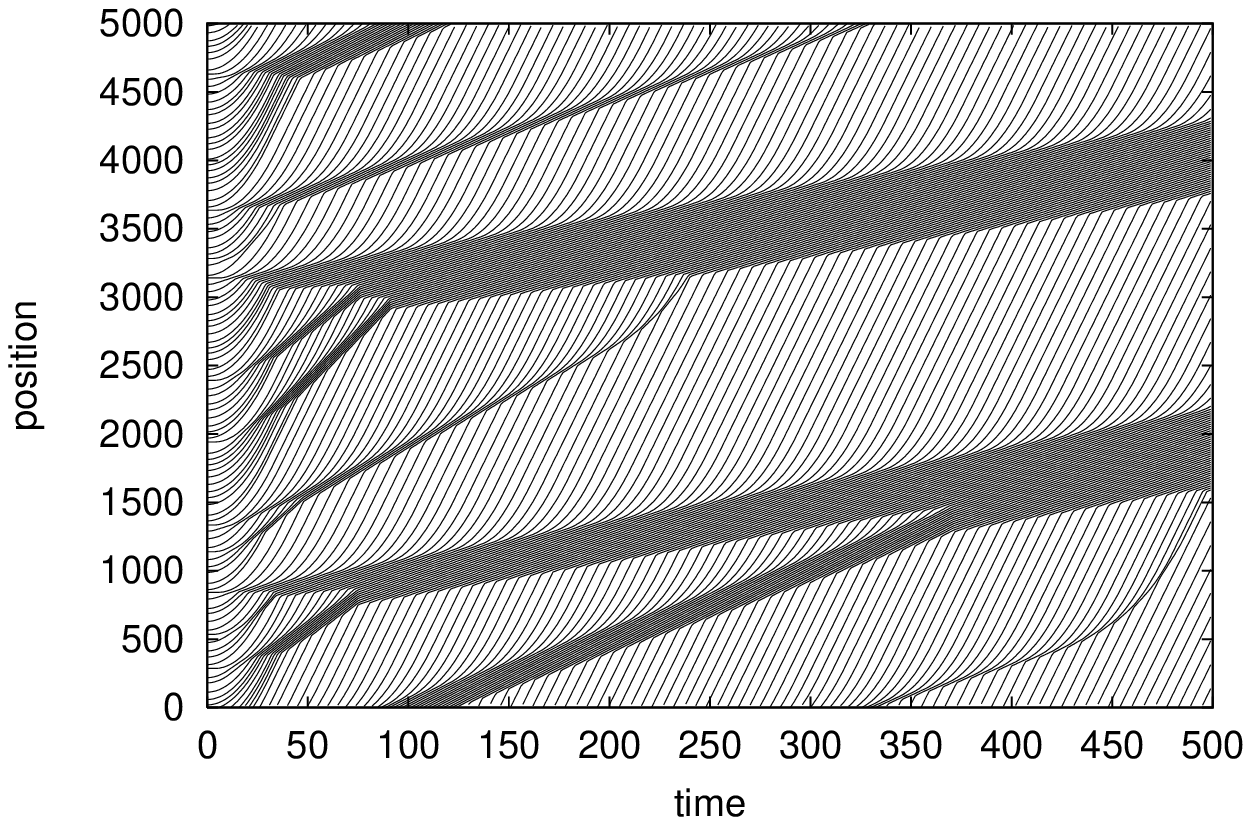}
  }
  \parbox{6cm}
  {
    \centering
    \psfrag{simulation}[r][r]{\tiny Simulation}
    \psfrag{average}[r][r]{\tiny Average}
    \psfrag{density}[t][]{Density $\rho$ in $vehicles/km$}
    \psfrag{flux}[][]{Flux $\Phi$ in $vehicles/h$}
    \includegraphics[width=5.5cm]{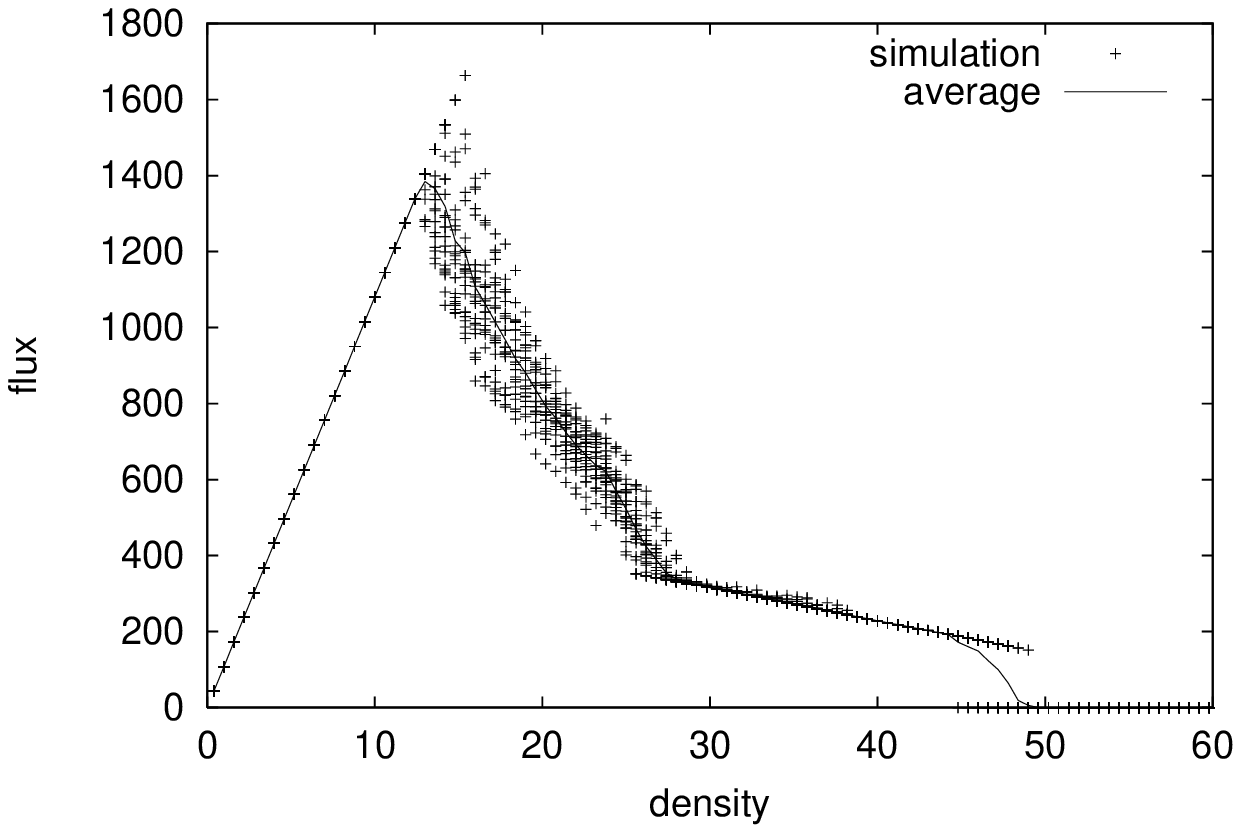}
  }
  \caption{The left side shows the trajectories of vehicles and jams
    with a velocity greater than 0 and a density of 20/km. The
    fundamental diagram is shown on the right side.
    Parameters: $\alpha=\delta=15, \beta=35,
    \gamma=10, a_{max}=1, v_{max}=30, p=0.75$}
\end{figure}

Equidistant positions including small fluctuations are used for the
initial configuration.
\begin{equation}
  x_i(t=0) = \frac{i}{\rho} + \frac{p}{2 \rho}
\end{equation}
The random number $p$ generates the fluctuations and characterizes their
magnitude with respect to the density. 

A higher maximum flux and critical density are obtained compared to
the homogeneous random case.
From the critical density up to the
maximum equidistant density $1/\beta$, we get congested areas consisting
of vehicles with a velocity $v_i>0$ that leads to a greater mean
velocity and a greater flux. From the equidistant maximum density
$1/\beta$ up to a cutoff density $\rho_{cut}$, the flux matches exactly
the homogeneous random case. The cutoff density
$\rho_{cut}$ can be calculated directly from the acceleration
condition including the fluctuations.
\begin{equation}
  \rho_{cut} = \frac{1+p}{\beta}
\end{equation}
%

\section{Outlook}

A simple 1d traffic model with deterministic propagation rules has
been studied. A strong dependence on the fluctuations of the initial
state has been found and for the extreme cases analytical solutions
can be obtained for the fundamental diagrams.

It would be interesting to generalize the present model to more than one
lane and to introduce more realistic velocity dependent threshold
parameters. In the vicinity of the critical density investigations
with respect to bistable states would be highly interesting.



\end{document}